\documentclass[10pt,letterpaper]{article}

\usepackage{ulem}
\usepackage[left=0.76in,right=0.76in,top=0.62in,bottom=0.66in]{geometry}
\usepackage{amsmath,amssymb,bm,mathtools}
\usepackage{graphicx}
\usepackage{xcolor} 
\usepackage{booktabs}
\usepackage{array}
\usepackage{pgfplots}
\usepackage{hyperref}
\usepackage{authblk}
\usepackage{titlesec}
\usepackage{caption}

\newcommand{\newc}{\newcommand}
\newc{\EV}{~{\rm eV}}

\pgfplotsset{compat=1.18}

\renewcommand{\thesection}{\Roman{section}}
\renewcommand{\thesubsection}{\Alph{subsection}}
\titleformat{\section}{\normalfont\normalsize\bfseries}{\thesection}{0.6em}{\MakeUppercase}
\titleformat{\subsection}{\normalfont\normalsize\bfseries}{\thesubsection}{0.5em}{}
\titlespacing*{\section}{0pt}{9pt}{4pt}
\titlespacing*{\subsection}{0pt}{7pt}{3pt}
\title{\vspace{-1.0em}\bfseries Black hole/quantum machine learning correspondence\vspace{-0.4em}}
\author[1]{Zae Young Kim\thanks{josephray@spinormedia.com}}
\author[2]{Jae-Weon Lee\thanks{scikid@jwu.ac.kr}}
\affil[1]{Spinor Media Inc, Seoul, 07332, Republic of Korea}
\affil[2]{Department of Electrical and Electronic Engineering, Jungwon University, 85 Munmu-ro, Goesan-eup, Goesan-gun, Chungcheongbuk-do, 28024, Republic of Korea}
\date{}

\begin{document}
\maketitle
\vspace{-1.3em}
\begin{abstract}

We explore a possible connection between the black hole information paradox and 
interpolation geometry underlying the double descent phenomenon in quantum machine learning. 
State-dependent operator reconstruction on the Hawking radiation can be formulated as
 a quantum linear inverse problem defined on the black hole-radiation purification. In this picture, the Page time corresponds to the interpolation threshold, where the dimensions of the remaining black hole and the radiation become comparable. 
 Using the Marchenko-Pastur law, we study the spectrum of the corresponding Gram matrices and obtain the variance of the linear reconstruction coefficients. The Page point is then related to both a change in the rank structure of the two subsystems and a strong enhancement of the coefficient variance near the interpolation threshold. This suggests a possible relation between the Page transition and interpolation phenomena in quantum machine learning.

\end{abstract}

\section{Introduction}

The black hole (BH) information problem is a paradox which originates from a contradiction between the {semiclassical} evaporation of BHs and {unitarity} in quantum mechanics~\cite{Hawking1976,Almheiri2021}. 
For an initially pure, closed system undergoing unitary evaporation, {the joint state of the remaining black hole and the Hawking radiation remains pure. Under the additional typicality assumption adopted here, the radiation entropy is approximately described by the Page mean}~\cite{Page1993a,Page1993b}. {We identify the dimension-balance point $N=P$ with the Page point within the adopted model. 
Here, $N$ and $P$ denote the black hole and
radiation dimensions, respectively.
For a generic purification, the dimension-balance point also marks the change in which subsystem has full Schmidt support.}

Operator reconstruction can be viewed as the problem of representing an action on one subsystem by an operator acting on another.
In holographic quantum error correction (QEC) and black hole models, state-specific reconstruction is often characterized by the norm of the difference between the states produced by two operator actions~\cite{Akers2022,Kar2023,Akers:2022qdl,
Antonini:2024yif}.
We describe the black hole-radiation purification by a coefficient matrix
$D\in\mathbb C^{N\times P}$.
The columns of $D$ contain the black hole amplitudes for the radiation basis states. For a fixed black hole operator $W_B$, the reconstruction condition is
\begin{equation}
W_BD=DW_R^T.
\end{equation}
For fixed $D$ and $W_B$, {finding an exact or least-squares radiation representative $W_R$ is a linear inverse problem.}
$D$ acts as the design matrix, $W_R^T$ contains the adjustable coefficients, and $W_BD$ is the target matrix in this problem.
The corresponding parameter-to-constraint ratio is $P/N$, and the
minimum reconstruction error follows from the Gram operators
$D^\dagger D$ and $DD^\dagger$.

A similar interpolation geometry appears in high-dimensional learning systems. In linear regression and some modern machine learning models,
the variance of an {unregularized} estimator can be large when the number of coefficients {approaches} the number of independent constraints.
This defines the interpolation threshold. Beyond this point, the prediction error of a minimum-norm interpolating solution can decrease again, giving the double-descent profile~\cite{Belkin2019,Loog2020,Schaeffer2023}.
For isotropic random features, the peak can be traced to eigenvalues of the design Gram matrix lying close to zero. The eigenvalue distribution can be described by the Marchenko-Pastur (MP) law ~\cite{Marchenko1967,speicher2023hda}.
The same spectral mechanism can also appear in quantum feature models and quantum kernels~\cite{Schuld:2018uel,Kempkes:2025hiw,Tomasi:2025wkm,Kamisoyama:2026psl}.


The central mathematical object is the coefficient matrix of a purified black hole-radiation state. We write
\begin{equation}
N\equiv \dim\mathcal H_B=e^S,\qquad P\equiv \dim\mathcal H_R=\Omega.
\label{eq:dimensions}
\end{equation}

There are two dimensional scales in the construction. The black hole side is characterized by the Bekenstein-Hawking entropy $S$, from which we take $N=e^S$ as an effective microcanonical dimension, while the finite radiation sector has dimension $P=\Omega$. This is a modeling choice, not a result that follows from the reconstruction equation.
After specifying the Hilbert-space dimensions and choosing the purification, the rank and geometry of the coefficient matrix control the reconstruction, while $\alpha\equiv P/N$ naturally characterizes the relative dimensions of the two sectors.

 The purification matrix has $N$ rows and $P$ columns, with aspect ratio
\begin{equation}
\alpha=\frac{P}{N}=\frac{\Omega}{e^S}.
\label{eq:alpha}
\end{equation}

For a specified black hole observable $W_B$ and radiation representative $W_R$, the purification defines the state-dependent reconstruction error $\|(W_B\otimes I_R-I_B\otimes W_R)|\psi\rangle\|^2$. In the coefficient representation, the reconstruction error becomes $P^{-1}\|W_BD-DW_R^T\|_F^2$. For each column, the reconstruction equation is an $N\times P$ linear system with $N$ black hole component conditions and $P$ adjustable radiation amplitudes. The resulting interpolation aspect ratio is therefore $P/N$. The row vectors of $D$ serve as $P$-component quantum features, while $W_BD$ specifies the corresponding target action on the purified state.

The reduced density matrices are determined by the two Gram
matrices of the same coefficient matrix, with a transpose on
the radiation side in the chosen basis. Therefore, the same coefficient geometry governs the Page curve and the MP spectrum, while also controlling the reconstruction error. Here, $N$ counts black hole basis components, not independently
prepared training examples. The columns of $Y=W_BD$ specify
the action of a prescribed black hole observable on a fixed
purification. They are not externally assigned labels for
unknown microstates. The learning interpretation concerns the
least-squares structure of this inverse problem. It does not define a supervised learning protocol for identifying an unknown black hole state from radiation measurements.

The Page point has two roles in this formulation. First, it is a rank and interpolation threshold. For $P<N$, the black hole reduced state has a $P$-dimensional Schmidt support $\mathcal H_B^\psi\subsetneq\mathcal H_B$, and a generic operator action need not preserve that support. The corresponding minimum reconstruction error is then generically nonzero. For $P>N$, a generic $D$ has full row rank, so $\mathcal H_B^\psi=\mathcal H_B$ and an exact state-dependent general linear radiation representative can be constructed for a specified black hole operator on the given purified state. A Hermitian radiation observable obeys the additional compatibility condition discussed below. Second, it is a conditioning threshold. The lower edge of the positive MP spectrum closes at $\alpha=1$, so finite statistical errors in the {target amplitudes} are strongly amplified by inverse Gram eigenvalues. The divergence therefore belongs to the estimator variance, not to the noiseless minimum reconstruction error itself.

The construction describes the kinematics of typical
purifications and the conditioning of a state-dependent
inverse problem. Where we invoke the double-descent analogy,
it refers only to the two-sided interpolation peak of the
coefficient variance. We do not identify this variance with
out-of-sample prediction risk, the full prediction risk,
or a radiation-decoding error. A state-independent decoder,
a protocol for accessing the target amplitudes, and a
dynamical quantum phase transition require additional
physical input.

The MP law and its inverse-moment relations are standard random-matrix results.
In this work, we use the MP law to analyze a purification-induced inverse problem, with particular attention to support leakage, the Hermitian compatibility condition, and coefficient sensitivity.

{In Sec. II we introduce the purification matrix and review its Page-rank and spectral properties. In particular, the Gram operators of the purification matrix determine the reduced density matrices and, under Haar-typicality, obey the MP law in the large-dimensional limit, with the Page point corresponding to the rank threshold $\alpha=P/N=1$. In Sec. III we use these geometric and spectral properties of the same purification matrix to derive the state-dependent operator-reconstruction error and its least-squares form. Its column-space geometry determines the rank-dependent minimum reconstruction error, while inverse eigenvalue moments of its Gram operators determine the fluctuation variance,
which we evaluate using the MP law.}
Section~IV discusses the physical content and limitations.

\section{The Page curve and the MP distribution}

A black hole with horizon area $A$ has Bekenstein-Hawking entropy
~\cite{Hawking1975}
\begin{equation}
S=\frac{A}{4G}.
\end{equation}
We use units $\hbar=c=k_B=1$.
Using the subsystem dimensions in Eq.~\eqref{eq:dimensions}, a pure state on $\mathcal H_B\otimes\mathcal H_R$ can be written as
\begin{equation}
|\psi\rangle=\frac{1}{\sqrt P}\sum_{i=1}^{P}|\psi_i\rangle\otimes|i\rangle,
\end{equation}
where $\{|i\rangle\}$ is an orthonormal radiation basis and the black hole vectors $|\psi_i\rangle\in\mathcal H_B$ are generally neither orthogonal nor individually normalized. 
{The vector $|\psi_i\rangle$ may be viewed as the black hole state associated with the radiation basis outcome $|i\rangle$. Thus, although the full black hole Hilbert space has dimension $N$, the particular purification $|\psi\rangle$ only involves the subspace spanned by the vectors ${|\psi_i\rangle}$.}
{Their expansion in an orthonormal basis $\{|a\rangle\}_{a=1}^{N}$ is}
\begin{equation}
|\psi_i\rangle=\sum_{a=1}^{N}D_{ai}|a\rangle,\qquad D\in\mathbb C^{N\times P}{.}
\end{equation}
{Normalization} of $|\psi\rangle$ gives
\begin{equation}
\frac{1}{P}\operatorname{Tr}(D^\dagger D)=1.
\end{equation}
{Equivalently, the $i$-th column of $D$ is the coordinate vector of $|\psi_i\rangle$ in the black hole basis. Hence the column space $\operatorname{Col}(D)$ is precisely the black hole subspace explored by the given purification.}


The reduced density matrices are

\begin{equation}
\rho_R=\frac{1}{P}D^T D^*
      =\frac{1}{P}(D^\dagger D)^T,
\qquad
\rho_B=\frac{1}{P}DD^\dagger.
\end{equation}
Here, the transpose is taken in the radiation basis used in the
purification. Indeed,
\begin{equation}
(\rho_R)_{ij}=\frac{1}{P}\sum_a D_{ai}D_{aj}^*. 
\end{equation}
Thus $D^\dagger D=P\rho_R^T$, whereas $DD^\dagger=P\rho_B$.
The two Gram matrices $D^\dagger D$ and $DD^\dagger$ have the same
nonzero eigenvalues. Since transposition preserves eigenvalues,
$\rho_R$ and $\rho_B$ therefore have the same nonzero spectrum. Similar Gram matrices occur in gravitational calculations of black hole microstate overlaps and replica geometries~\cite{Penington:2019kki,Kawabata:2021hac,Balasubramanian:2022gmo,Balasubramanian:2022lnw,Climent:2024trz,Muck:2024fpb,Iizuka:2024njd}.

{The second relation follows by tracing the pure state over the radiation basis: the radiation trace identifies the radiation indices of $D$ and $D^\dagger$, yielding $DD^\dagger/P$ as the reduced density operator on $\mathcal H_B$. Thus $DD^\dagger$ is not merely a Gram matrix. Up to the normalization factor $P$, it is the physical reduced density matrix of the remaining black hole.
}

We model a typical purification by a Haar-random pure state
on the chosen product space. Equivalently, let
$G\in\mathbb{C}^{N\times P}$ have independent standard circular
complex Gaussian entries, with $\mathbb{E}|G_{ai}|^2=1$, and set
\begin{equation}
D=\frac{\sqrt{P}\,G}{\|G\|_F}.
\end{equation}
Here, standard circular complex Gaussian means that one may write
$G_{ai}=x_{ai}+iy_{ai}$ with independent real Gaussian variables
$x_{ai}$ and $y_{ai}$ of zero mean and variance $1/2$. Hence
$\mathbb{E}|G_{ai}|^2=1$, and the distribution does not prefer any
complex phase direction. This construction is a convenient way to
sample a Haar-random pure state after normalization.
This enforces $\operatorname{Tr}(D^\dagger D)=P$ exactly.
The normalization correlates the entries at finite size.
At fixed aspect ratio, their Gram spectrum has the same
large-dimensional MP limit as that of $G/\sqrt{N}$.
Haar typicality is an ensemble assumption, not a consequence
of the reconstruction equation or of unitarity alone.

The reconstruction identities do not require Haar randomness.
{For any specified $D$ satisfying the stated rank assumptions, the reconstruction equation, the Schmidt-support projector, and the minimum leakage formula for a general linear solution follow from linear algebra. The maximal-rank relation $\operatorname{rank}D=\min(N,P)$ holds for a generic purification, and formulas involving ordinary Gram-matrix inverses require the corresponding full-rank assumptions.}
The Hermiticity
condition $[W_B,\rho_B]=0$ is also a fixed-$D$ statement. In the
full-row-rank regime it is necessary and sufficient for the existence
of an exact Hermitian radiation representative. Haar typicality is used
only for ensemble quantities such as the Page-mean entropy, the MP
spectrum, and the asymptotic inverse-Gram variance.

 In the limit
\begin{equation}
N,P\to\infty,\qquad \alpha=\frac{P}{N}=\frac{\Omega}{e^S}\quad\text{fixed},
\end{equation}
the empirical eigenvalue density of {$D^\dagger D=P\rho_R^T$} converges to the MP law
\begin{align}
f_{\mathrm{MP}}(\lambda)
&=\frac{\sqrt{(\lambda-\lambda_-)(\lambda_+-\lambda)}}{2\pi\alpha\lambda}
\Theta(\lambda-\lambda_-)\Theta(\lambda_+-\lambda)\nonumber\\
&\quad+\left(1-\frac{1}{\alpha}\right)\Theta(\alpha-1)\,\delta(\lambda),
\end{align}
with
\begin{equation}
\lambda_\pm=(1\pm\sqrt\alpha)^2.
\end{equation}
For $\alpha>1$,
a generic purification has $\operatorname{rank}\rho_R=N$, so
$\rho_R$ has $P-N$ zero eigenvalues because its rank cannot exceed $N$. These zero modes make up a fraction $(P-N)/P$ of the radiation space spectrum and appear as a delta function contribution at zero.
Conversely, for $\alpha<1$, $DD^\dagger=P\rho_B$ has $N-P$ zero modes. {The two Gram matrices have the same positive spectrum and therefore the same smallest positive eigenvalue, which controls the amplification associated with inverting their positive modes.}

The radiation entropy is
\begin{equation}
S_R=-\operatorname{Tr}(\rho_R\log\rho_R).
\end{equation}
Let $\overline S_R=\mathbb E_{\mathrm{Haar}}[S_R]$.
For $m,n\to\infty$ at fixed $m/n>0$, the Haar-mean entropy is
\begin{equation}
\overline S_R=\log m-\frac{m}{2n}+O(n^{-2}),
\qquad m=\min(P,N),\quad n=\max(P,N).
\end{equation}

Equivalently, at fixed $\alpha>0$,
\begin{equation}
\overline S_R=
\begin{cases}
S+\log\alpha-\alpha/2,&0<\alpha\le1,\\
S-1/(2\alpha),&\alpha\ge1,
\end{cases}
\quad +O(n^{-2}).
\end{equation}
At $\alpha=1$, this gives
$\overline S_R=S-1/2+O(e^{-2S})$.
These are ensemble-mean statements. An individual Haar-typical
state has an entropy concentrated near the mean rather than
an exactly fixed entropy. 
Exchanging $N$ and $P$ leaves the entropy unchanged and sends
$\alpha$ to $1/\alpha$. {Under this subsystem interchange, the reduced density matrices exchange their roles and retain the same nonzero spectrum.}

\section{State-dependent reconstruction and spectral conditioning}

\subsection{Purification and state-dependent reconstruction}

{The  results derived below probe different aspects of the same purification matrix introduced in Sec.~II. The minimum reconstruction error is controlled by the column-space geometry of $D$, whereas the fluctuation variance is controlled by inverse eigenvalue moments of its Gram operators $D^\dagger D$ and $DD^\dagger$. Thus, the Page-rank transition and the variance enhancement near $\alpha=1$ originate respectively from the rank and spectral properties of the same purification matrix.}
At a fixed evaporation time, the matrix $D$ contains the amplitudes of the purified black hole-radiation state. Its $i$-th column $d_i=(D_{1i},\ldots,D_{Ni})^T$ is the unnormalized black hole vector conditioned on the radiation basis state $|i\rangle$. Hence $D^\dagger D=P\rho_R^T$ is the Gram matrix of these conditional black hole states, with $(D^\dagger D)_{ij}=d_i^\dagger d_j$.

For the Quantum machine learning (QML) interpretation, write the $a$-th row as the $P$-component vector $\phi_a^\dagger=(D_{a1},\ldots,D_{aP})$. Then $(DD^\dagger)_{ab}=\phi_a^\dagger\phi_b=P(\rho_B)_{ab}$. Thus the rows of $D$ have the same algebraic role as feature vectors in the associated linear inverse problem, although the index $a$ labels black hole basis components rather than independently prepared training samples.

For a generic purification,
\begin{equation}
\operatorname{rank}D=\operatorname{rank}\rho_R=\operatorname{rank}\rho_B=\min(N,P).
\end{equation}
We denote the black hole Schmidt support by
\begin{equation}
\mathcal H_B^\psi\equiv\operatorname{supp}(\rho_B)\subseteq\mathcal H_B,
\qquad \dim\mathcal H_B^\psi=\operatorname{rank}D,
\end{equation}
and write $\Pi_B^\psi$ for the orthogonal projector onto this physically occupied black hole sector.
Because $\rho_B=DD^\dagger/P$, its support is exactly the column space of $D$, so that $\mathcal H_B^\psi=
\operatorname{supp}(\rho_B)=\operatorname{Col}(D)$. 
Physically, $\mathcal H_B^\psi$ is the part of the black hole Hilbert space that actually participates in the entanglement with the radiation for the particular state $|\psi\rangle$. Directions orthogonal to this support are absent from the purification.
Let $W_B=W_B^\dagger$ be an observable acting on the remaining black hole. We ask whether its action on the purified state can be represented by an operator $W_R$ acting only on the radiation. Here $B$ and $R$ are subsystem labels, while $a,b,j,\mu$ denote basis indices. The state-dependent condition is
\begin{equation}
\label{condition}
(W_B\otimes I_R)|\psi\rangle=(I_B\otimes W_R)|\psi\rangle.
\end{equation}
{In words, Eq. (\ref{condition}) asks whether the change in the joint state produced by acting on the black hole can be reproduced exactly by acting on the radiation alone. This is a state-dependent statement: $W_R$ is required to reproduce the action of $W_B$ on the particular entangled state $|\psi\rangle$, not necessarily on arbitrary states in $\mathcal H_B\otimes\mathcal H_R$.}
Using the state expansion above, this condition becomes
$
W_BD=DW_R^T,
$
which is a general coefficient-matrix form of state-dependent operator transfer across a bipartite purification.
The transpose appears because of the way the coefficient matrix is defined.
This relation is state dependent and does not constitute an operator
identity between the two subsystem Hilbert spaces.

{Every column of $DW_R^T$ is a linear combination of the columns of $D$ and therefore necessarily lies in $\operatorname{Col}(D)=\mathcal H_B^\psi$. By contrast, $W_BD$ need not lie entirely in this subspace, because $W_B$ can rotate a vector in $\mathcal H_B^\psi$ into a direction orthogonal to $\mathcal H_B^\psi$. Exact reconstruction is possible only when this orthogonal component vanishes.}
In components, the same condition is $\sum_{b=1}^{N}(W_B)_{ab}D_{b\mu}=\sum_{j=1}^{P}D_{aj}(W_R)_{\mu j}$ for $a=1,\ldots,N$ and $\mu=1,\ldots,P$.

To quantify imperfect reconstruction, define the state-dependent residual
\begin{equation}
|\Delta(W_R)\rangle\equiv(W_B\otimes I_R-I_B\otimes W_R)|\psi\rangle.
\end{equation}
Norms of differences between the states generated by the two operator actions provide standard diagnostics of approximate or state-specific reconstruction in holographic QEC~\cite{Akers2022,Kar2023,Akers:2022qdl}. Ref.~\cite{Antonini:2024yif} used an averaged squared reconstruction error, minimized over unitary representatives, to quantify state dependence. Here the state is fixed and $W_R$ is a general linear representative. Expanding in the product basis gives
\begin{equation}
|\Delta(W_R)\rangle=\frac{1}{\sqrt P}\sum_{a=1}^{N}\sum_{\mu=1}^{P}
(W_BD-DW_R^T)_{a\mu}|a,\mu\rangle.
\end{equation}
With the orthonormal product basis, the state norm reduces to the Frobenius norm of the coefficient residual,
\begin{equation}
L(W_R)\equiv\|\Delta(W_R)\|^2=\frac{1}{P}\|W_BD-DW_R^T\|_F^2.
\end{equation}
{Accordingly, $L(W_R)$ measures the squared Hilbert-space distance between the state produced by the black hole action and the state produced by its radiation-side representative. It therefore has a direct state-reconstruction meaning rather than being introduced merely as a formal regression loss.}
The Hilbert-space geometry therefore induces a least-squares functional for the purification coefficients.
Writing $Y\equiv W_BD$, $y_\mu\equiv Y_{:\mu}\in\mathbb C^N$, and
$w_\mu\equiv(W_R^T)_{:\mu}\in\mathbb C^P$, we obtain
\begin{equation}
r_\mu\equiv y_\mu-Dw_\mu,\qquad
L(W_R)=\frac{1}{P}\sum_{\mu=1}^{P}\|r_\mu\|_2^2.
\end{equation}
Using the row vectors above, the reconstructed component is $(Dw_\mu)_a=\phi_a^\dagger w_\mu$. Thus, for each fixed $\mu$, the problem is an $N$-component linear reconstruction with $P$ adjustable amplitudes. 
The corresponding interpolation ratio is
\begin{equation}
\frac{\text{adjustable amplitudes}}{\text{component constraints}}=\frac{P}{N}=\alpha.
\end{equation}
Summing the component count over all $\mu$ gives $NP$ equations for the $P^2$ entries of $W_R$. Hence the number of unknowns relative to constraints is again $P^2/(NP)=P/N$.
Thus, the dimensions of the two subsystems fix the rectangular interpolation geometry.

\begin{figure*}[tb]
\centering
\includegraphics[width=0.8\textwidth]{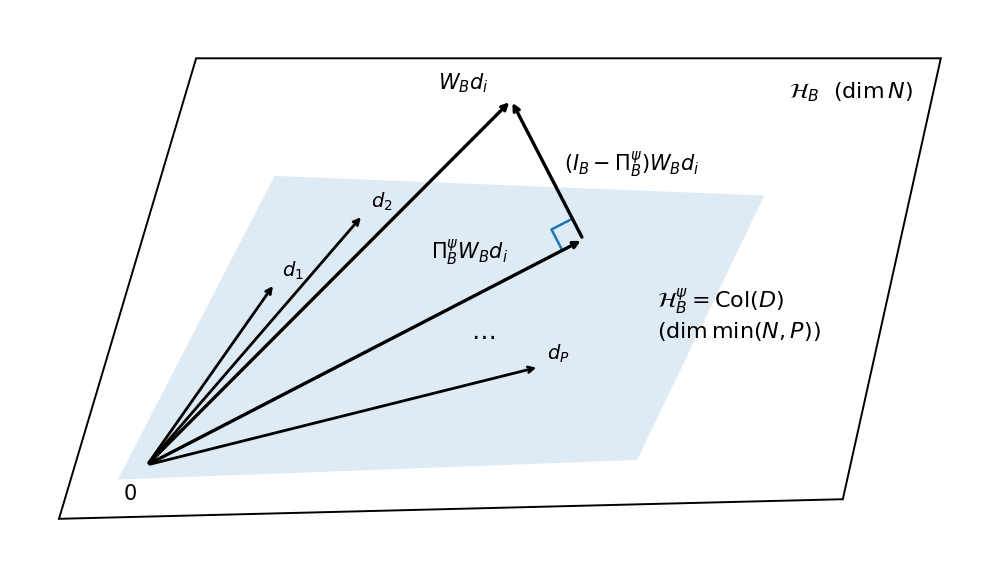}
 \caption{ Geometric interpretation of state-dependent reconstruction in the black hole Hilbert space $\mathcal{H}_B$. The column vectors $d_i$ of the purification matrix $D$ span the Schmidt support $\mathcal{H}_B^\psi=\mathrm{Col}(D)$. The action $W_B d_i$ is decomposed into the component $\Pi_B^\psi W_B d_i$ within the Schmidt support and the orthogonal component $(I_B-\Pi_B^\psi)W_B d_i$ outside it. The latter cannot be reproduced by any radiation-side action $D W_R^T$ and therefore represents the irreducible reconstruction error before the Page point.
}
 \label{lambda}
\end{figure*}

Equation~(\ref{condition}) is a stronger statement than matching one expectation value. It requires that, for every radiation operator $X_R$, $\langle\psi|W_B\otimes X_R|\psi\rangle=\langle\psi|I_B\otimes W_R^\dagger X_R|\psi\rangle$.
The radiation representative is intrinsically state dependent.
Stationarity of the least-squares functional with respect to the complex amplitudes gives the normal equation
\begin{equation}
D^\dagger(DW_R^T-W_BD)=0.
\end{equation}
This equation is sufficient to derive the reconstruction threshold directly from the rank of the two Gram operators.

\subsection{Page threshold and reconstruction}

Before the Page time, $P<N$ and a typical $D$ has full column rank. The black hole Schmidt support has dimension $P$ and its projector is
\begin{equation}
\Pi_B^\psi=D(D^\dagger D)^{-1}D^\dagger.
\end{equation}
The formula follows entirely from the positive-definite Gram matrix $D^\dagger D$: $\Pi_B^\psi$ is Hermitian, $(\Pi_B^\psi)^2=\Pi_B^\psi$, and $\Pi_B^\psi D=D$. Hence it projects onto the black hole sector occupied by the purification.
{For any black hole vector $|v\rangle$, $\Pi_B^\psi|v\rangle$ retains the component that can be expressed as a linear combination of the black hole states already present in the purification, whereas $(I_B-\Pi_B^\psi)|v\rangle$ extracts the component outside that Schmidt support. (See Fig. 1.)}

The normal equation has the unique least-squares solution
\begin{equation}
W_{R,<}^T=(D^\dagger D)^{-1}D^\dagger W_BD.
\end{equation}
Its reconstructed target is $DW_{R,<}^T=\Pi_B^\psi W_BD$, so the minimum residual and reconstruction error are
\begin{equation}
R_<=(I_B-\Pi_B^\psi)W_BD,\qquad
L_{\min}^{<}=\frac{1}{P}\|R_<\|_F^2.
\end{equation}

This expression makes the origin of the pre-Page reconstruction error clear.
The radiation-side action $DW_R^T$ can reproduce the projection $\Pi_B^\psi W_BD$, but it cannot reproduce the orthogonal component $(I_B-\Pi_B^\psi)W_BD$, because multiplication by $D$ can generate only vectors in $\operatorname{Col}(D)$. The residual is therefore a geometric obstruction imposed by the Schmidt support, rather than an error arising from an imperfect choice of $W_R$.

Using $DD^\dagger=P\rho_B$, the same quantity is
\begin{equation}
L_{\min}^{<}=\operatorname{Tr}\!\left[\rho_BW_B^\dagger(I_B-\Pi_B^\psi)W_B\right].
\end{equation}
Writing $\rho_B=\sum_{k=1}^{P}p_k|u_k\rangle\langle u_k|$, one obtains
\begin{equation}
L_{\min}^{<}=\sum_{k=1}^{P}p_k\|(I_B-\Pi_B^\psi)W_B|u_k\rangle\|^2.
\label{rev:eq:schmidt-leakage}
\end{equation}
Thus $L_{\min}^{<}$ is precisely the Schmidt-weighted leakage of the black hole action from the physically occupied black hole sector into its orthogonal complement.
A vanishing residual occurs if and only if
\begin{equation}
(I_B-\Pi_B^\psi)W_B\Pi_B^\psi=0,
\end{equation}
{which states that $W_B$ must map every black hole state participating in the purification back into the same Schmidt support. Before the Page time this is a nontrivial restriction, because $\mathcal H_B^\psi$ is only a proper subspace of $\mathcal H_B$.}
A generic $W_B$ does not satisfy this condition. In the associated QML problem, $P<N$ is the underparameterized side of the interpolation threshold, and $R_<$ is a deterministic geometric obstruction rather than a statistical fluctuation.

After the Page time, $P>N$ and a typical $D$ has full row rank. Then $DD^\dagger$ is positive definite and one may define directly
\begin{equation}
W_{R,>}^T=D^\dagger(DD^\dagger)^{-1}W_BD.
\end{equation}
Multiplication by $D$ gives
\begin{equation}
DW_{R,>}^T=DD^\dagger(DD^\dagger)^{-1}W_BD=W_BD,
\qquad L_{\min}^{>}=0.
\end{equation}
{The geometric reason for this qualitative change is simple. For $P>N$, a generic $N\times P$ matrix $D$ has rank $N$, and hence $\operatorname{Col}(D)$ has the same dimension as $\mathcal H_B$. Therefore $\operatorname{Col}(D)=\mathcal H_B$ and $\mathcal H_B^\psi=\mathcal H_B$: there is no longer any black hole direction outside the Schmidt support into which $W_B$ can leak.
Consequently, the obstruction represented by $I_B-\Pi_B^\psi$ disappears, and every target column of $W_BD$ can be generated by a suitable linear combination of the columns of $D$.}

When $P=N$, the subsystem dimensions coincide and $\alpha=1$.
A generic finite square $D$ is invertible, so one can see that
$W_R^T=D^{-1}W_BD$ and $L_{\min}=0$.
These exact solutions are obtained over unrestricted linear radiation
operators.
This is weaker than reconstructing a physical radiation
observable.

A physical observable must be Hermitian, while the linear
solution above need not be Hermitian even when $W_B$ is Hermitian.
In the full-row-rank regime an exact Hermitian radiation representative
exists if and only if
\begin{equation}
[W_B,\rho_B]=0.
\end{equation}
Thus $L_{\min}=0$ means that the state action can be reproduced by some
linear radiation operator. It does not mean that every black hole
observable has an exact radiation-observable representative. The
commutator condition is therefore a physically important restriction,
not merely a technical refinement.

Two distinct questions should be separated at this point. Solvability
asks whether a radiation-side representative exists at all, whereas
conditioning asks how sensitively that representative changes under a
small perturbation of the target. A linear system may have an exact
solution and nevertheless be extremely ill conditioned. In the present
problem the rank obstruction disappears when $D$ has full row rank, but
the inverse problem becomes most sensitive when its smallest positive
singular value approaches zero.

The sharp large-dimensional behavior instead arises because the lower MP edge
\begin{equation}
\lambda_-=(1-\sqrt\alpha)^2
\end{equation}
closes at $\alpha=1$. 
{As $\alpha$ approaches unity, $\lambda_-$ approaches zero, so the Gram matrix develops increasingly weakly constrained directions. Inverting the Gram operator then magnifies small errors along these directions, explaining why reconstruction can become statistically unstable even when an exact noiseless solution exists.}
Across the same rank threshold, the reconstruction changes from generic Schmidt-support leakage to a regime admitting exact state-dependent representation, while the inverse problem becomes increasingly sensitive near the threshold. A single radiation operator valid for a family of purifications would have to satisfy the stronger condition $W_BD_\beta=D_\beta W_R^T$ for every $\beta$, which is not guaranteed by the dimension count or by the fixed-state minimum alone.
{The purification matrix $D\in\mathbb{C}^{N\times P}$ has two complementary interpretations. Its columns $d_i\in\mathcal{H}_B$ span the Schmidt support $\mathcal{H}_B^\psi=\mathrm{Col}(D)$ and determine the reconstruction geometry,
whereas in the associated least-squares problem $y_\mu\approx Dw_\mu$, its rows
 act as $P$-component quantum feature vectors. Accordingly, $\alpha=P/N$ plays the role of the parameter-to-constraint ratio. Thus, black hole reconstruction and linear learning arise from the column and row geometries of the same purification matrix.
}

\subsection{Spectral fluctuations and Schwarzschild parametrization}

We now study the coefficient sensitivity of the induced
least-squares problem. Throughout this calculation, $D$ and
the prescribed observable $W_B$ are treated as known.
The matrix $Y=W_BD$ contains target amplitudes in the chosen
bases, rather than directly measured correlation functions.
We introduce an additive perturbation of this target and
assume the following conditional first and second moments:

\begin{equation}
\label{YE}
\widetilde Y=Y+E,\qquad
\mathbb E[E_{:\mu}\mid D]=0,\qquad
\mathbb E[E_{:\mu}E_{:\nu}^\dagger\mid D]=\sigma^2\delta_{\mu\nu}I_N,
\end{equation}

These moment assumptions define an isotropic target-noise
model. Gaussianity is not required for the variance calculation.
The model is used to diagnose sensitivity to target perturbations,
not to derive a measurement protocol for $Y$. In particular,
the relation between $\sigma^2$ and
 the available measurement resources depends
on the observables, state access, and sampling procedure, none
of which is specified here. Errors in $D$ and their correlations
with target errors are also outside the present calculation.
The resulting variance is therefore conditional on the adopted
target-noise model.

The perturbation $E$ should therefore be understood as a probe of the
local sensitivity of the inverse map, rather than as a microscopic model
of Hawking-radiation noise. Its role is analogous to applying small
perturbations in target-space directions and asking which directions are
most strongly amplified by inversion.

\begin{figure}[t]
\centering
\begin{tikzpicture}
\begin{axis}[
width=0.72\textwidth,
height=0.42\textwidth,
xmin=0,xmax=4,
ymin=-3,ymax=5,
xlabel={$\alpha=P/N=\Omega/e^S$},
ylabel={$V(\alpha)/\sigma^2$ and {$\overline S_R-S$}},
legend style={at={(0.98,0.98)},anchor=north east},
grid=major,
samples=250,
]
\addplot[thick,domain=0.001:0.96] {x/(1-x)};
\addplot[thick,forget plot,domain=1.04:4] {1/(x-1)};
\addlegendentry{$V(\alpha)/\sigma^2$}
\addplot[dashed,thick,domain=0.01:1] {ln(x)-x/2};
\addplot[dashed,thick,forget plot,domain=1:4] {-1/(2*x)};
\addlegendentry{{$\overline S_R-S$}}
\addplot[dotted,thick,forget plot] coordinates {(1,-3) (1,5)};
\end{axis}
\end{tikzpicture}
\caption{{Leading Haar-mean entropy shift $\overline S_R-S$
and asymptotic coefficient variance $V(\alpha)/\sigma^2$
as functions of $\alpha=P/N$. The vertical line marks
the dimension balance $P=N$. The entropy shift is not
the radiation entropy plotted along an evaporation trajectory.
The variance curve describes the unregularized isotropic
target-noise model with known $D$. It is not the noisy
state-action loss. The latter has noise contribution
$\sigma^2\operatorname{rank}D$, as derived in the text.
The coefficient pole is regularized by ridge regression
or an explicit cutoff on small Gram eigenvalues.}}
\label{fig:page_variance}
\end{figure}

For $P<N$, the least-squares fluctuation of one output channel is
\begin{equation}
\label{w<}
\delta w_<=(D^\dagger D)^{-1}D^\dagger\epsilon,
\end{equation}
so that the inverse eigenvalue moment of $D^\dagger D$ controls the variance. For $P>N$, the minimum-norm fluctuation is
\begin{equation}
\label{w>}
\delta w_>=D^\dagger(DD^\dagger)^{-1}\epsilon,
\end{equation}
so the relevant inverse moment changes to that of $DD^\dagger$ as the rank structure is exchanged across $\alpha=1$. 
Using $\delta w=D^{+}\epsilon$, we obtain
$
\|\delta w\|_2^2
=
\epsilon^\dagger(D^{+})^\dagger D^{+}\epsilon.
$
Here, $D^+$ is the Moore-Penrose pseudoinverse.
Taking the conditional expectation over the noise and using the cyclic
property of the trace gives
\begin{equation}
\begin{aligned}
\mathbb{E}_{\epsilon}[\|\delta w\|_2^2\mid D]
&=
\mathbb{E}_{\epsilon}
[\epsilon^\dagger(D^{+})^\dagger D^{+}\epsilon\mid D]
\\
&=
\operatorname{Tr}\!\left[
(D^{+})^\dagger D^{+}
\mathbb{E}_{\epsilon}[\epsilon\epsilon^\dagger\mid D]
\right]
\\
&=
\operatorname{Tr}\!\left[
D^{+}\Sigma(D^{+})^\dagger
\right].
\end{aligned}
\end{equation}
Here, $D$ is held fixed when taking the conditional expectation, and
$\Sigma\equiv \mathbb{E}[\epsilon\epsilon^\dagger\mid D]$ describes the noise strength and correlations in the target space.
Thus, the amplification depends not only on the small singular values of
$D$ but also on how the noise covariance projects onto the corresponding
singular-vector directions. 
For the isotropic model specified in Eq. (\ref{YE}), \(\Sigma=\sigma^2 I_N\), the fluctuations in Eqs. (\ref{w<}) and (\ref{w>}) yield conditional mean-square coefficient errors of \(\sigma^2\operatorname{Tr}[(D^\dagger D)^{-1}]\) and \(\sigma^2\operatorname{Tr}[(DD^\dagger)^{-1}]\), respectively.

Define the coefficient variance per black hole component by
\begin{equation}
V_{N,P}\equiv\frac{1}{N}
\mathbb{E}_{D,E}\|\delta w_\mu\|_2^2.
\end{equation}
For $N,P\to\infty$ with $P/N\to\alpha\ne1$, write
$V(\alpha)=\lim V_{N,P}$. The MP inverse moments then give
the two branches displayed below.

{See Fig.~\ref{fig:page_variance}.}

\begin{equation}
V(\alpha)=
\begin{cases}
\displaystyle \sigma^2\frac{\alpha}{1-\alpha},&0<\alpha<1,\\[1.2ex]
\displaystyle \sigma^2\frac{1}{\alpha-1},&\alpha>1.
\end{cases}
\end{equation}
The derivation is given in Appendix~A.
{It is important that this divergence does not mean that the exact reconstruction error diverges at the Page point. Rather, it means that a small uncertainty in the inferred target can produce a large uncertainty in the reconstructed radiation operator because the inverse Gram problem is nearly singular.}
On the $\alpha>1$ side, $D^\dagger D$ has a zero-eigenvalue sector of weight $1-1/\alpha$ when normalized over its $P$ eigenvalues, whereas $DD^\dagger$ is normalized over its $N$ positive eigenvalues. The variance expression therefore uses the inverse moment of the positive $N\times N$ Gram operator on that side and contains no additional factor of $1/\alpha$.

The coefficient variance $V$ should also be distinguished from
the expected state-action loss of the noisy estimator.
The coefficient variance measures the sensitivity of the reconstructed operator coefficients to target perturbations, whereas the state-action loss measures the actual mismatch between the black-hole and radiation-side actions on the purified state.
For fixed,
known $D$, define the unconstrained minimum-norm estimator
\begin{equation}
\widehat W_R^T=D^+(Y+E),\qquad
\Pi_B^\psi=DD^+,\qquad r=\operatorname{rank}D.
\end{equation}
The residual relative to the noiseless target is
\begin{equation}
Y-D\widehat W_R^T=(I_B-\Pi_B^\psi)Y-\Pi_B^\psi E.
\end{equation}
The two terms are orthogonal in the Frobenius inner product.
Consequently, under the stated conditional noise covariance,
\begin{equation}
\mathbb{E}_E\!\left[L(\widehat W_R)\mid D\right]
=L_{\min}+\frac{1}{P}
 \mathbb{E}_E\!\left[\|\Pi_B^\psi E\|_F^2\mid D\right]
=L_{\min}+\sigma^2 r.
\label{rev:eq:noisy-state-loss}
\end{equation}
For a generic purification, dividing by $N$ gives the normalized noise contribution $[\mathbb{E}_E(L(\widehat W_R)\mid D)-L_{\min}]/N=\sigma^2\min(\alpha,1)$, which has no interpolation pole. The MP divergence therefore characterizes
sensitivity of the inferred coefficients, not the state-action
loss evaluated on the same purification. Prediction on other
states would require a separate choice of test states and
error metric.
The variance curve shows the sensitivity of the reconstructed coefficients to isotropic perturbations of the target amplitudes, assuming that \(D\) is known exactly and that no regularization is applied. 
Ridge regularization or truncation of small Gram eigenvalues replaces the divergence of the coefficient variance at \(\alpha=1\) by a finite crossover.

Before the Page point, two different effects are present. One is the deterministic error $L_{\min}^{<}$ from Schmidt-support leakage, and the other is the fluctuation of the reconstructed coefficients, which is controlled by inverse Gram eigenvalues.
The variance formula describes only the latter.
At $\alpha=1$, the inverse MP moment is singular in the infinite-dimensional, unregularized limit. Moreover, in the ideal complex-Gaussian ensemble the ensemble average of the unregularized inverse Gram matrix also diverges at exact square dimension, even though a single finite square realization is invertible with probability one. Ridge regularization, Gram-eigenvalue truncation, or departures from the idealized ensemble can replace this singular susceptibility by a finite crossover.

A simple Schwarzschild specialization illustrates how the abstract aspect ratio can be related to the black hole mass. In four dimensions and units $G=\hbar=c=k_B=1$, $S(M)=4\pi M^2$. Let $M_0$ and $S_0=4\pi M_0^2$ denote the initial mass and entropy. As an illustrative dimension assignment, take the effective product dimension to remain $NP=e^{S_0}$, so that $N=e^{S(M)}$ and $P=e^{S_0-S(M)}$. The relation $\log P=S_0-S(M)$ is an additional model assumption. It does not follow from unitarity alone and does not include the coarse-grained entropy production of real Hawking radiation. Thus
\begin{equation}
\alpha(M)=\exp[S_0-2S(M)]=\exp[4\pi(M_0^2-2M^2)].
\end{equation}
The Page condition $\alpha=1$ gives
\begin{equation}
M_{\mathrm{Page}}=\frac{M_0}{\sqrt2}.
\end{equation}
This specialization is only a mass parametrization of the adopted inverse-problem model. Greybody factors, entropy production, and departures from Haar typicality can modify the physical identification of the effective dimensions and spectrum.

\begin{table}[t]
\caption{{Purification-induced correspondence with a linear inverse
problem. Rank and exact-fit statements assume a generic
purification and unrestricted linear radiation representatives.
Hermitian representatives require the additional condition shown.}}
\label{tab:correspondence}
\centering
\begin{tabular}{lll}
\toprule
Concept & black hole purification & QML interpretation\\
\midrule
Matrix & $D\in\mathbb C^{N\times P}$ & $N\times P$ reconstruction map\\
Dimensions & $N=e^S,\ P=\Omega$ & $N$ constraints, $P$ amplitudes\\
Gram operators & $P\rho_B=DD^\dagger,\ {P\rho_R^T=D^\dagger D}$ & common positive spectrum\\
Error & squared state-action norm & $P^{-1}\|Y-DW_R^T\|_F^2$\\
Aspect ratio & $\alpha=P/N=\Omega/e^S$ & interpolation ratio\\
$\alpha<1$ & $\mathcal H_B^\psi\subsetneq\mathcal H_B$ & underparameterized, generic support leakage\\
$\alpha=1$ & equal dimensions & interpolation/conditioning threshold\\
$\alpha>1$ & $\mathcal H_B^\psi=\mathcal H_B$ & {exact fit, general linear representative}\\
{Hermitian ($P\ge N$)} & {$[W_B,\rho_B]=0$} & {additional compatibility condition}\\
\bottomrule
\end{tabular}
\end{table}

Table~\ref{tab:correspondence} summarizes the correspondence.
Different aspects of the same purification geometry can be captured by the Page entropy, the minimum reconstruction error, and the peak in coefficient variance. The entropy is determined by the Schmidt spectrum, $L_{\min}$ by leakage outside the black-hole Schmidt support, and the estimator variance by inverse moments of the positive Gram eigenvalues.

\section{Discussion}

The reconstruction problem separates two issues that should not be confused. Exact solvability is controlled by the rank of the purification matrix, whereas statistical stability is controlled by the small positive eigenvalues of its Gram operators. Thus, an exact state-dependent solution can exist while the corresponding inverse problem remains strongly ill conditioned near the Page point.

A further restriction appears when the radiation representative is required to be Hermitian, as appropriate for an observable. If both $W_B$ and $W_R$ are Hermitian and exact reconstruction holds, then
\begin{equation}
[W_B,DD^\dagger]=0
\qquad\Longleftrightarrow\qquad
[W_B,\rho_B]=0.
\end{equation}
In the full-row-rank regime this condition is also sufficient. Therefore, the existence of a general linear radiation representative does not imply that every black hole observable has an exact Hermitian representative on the radiation. 
For a nondegenerate $\rho_B$, this condition forbids $W_B$ from mixing Schmidt states with unequal weights. Approximate Hermitian reconstruction outside this compatible sector lies beyond the scope of the present analysis. The reconstruction studied here is state dependent.
 A single radiation operator acting on a code subspace would have to satisfy the reconstruction condition for a family of purifications rather than for one fixed state. Operational reconstruction would also require access to $D$ and the target amplitudes together with a measurement and estimation protocol. Radiation data alone do not determine the purification: $D$ and $U_BD$ give the same $\rho_R$ for any black hole unitary $U_B$, but generally require different representatives of a fixed $W_B$.

The Schwarzschild specialization is intended only as an illustrative mass parametrization. Its quantitative form assumes $P=e^{S_0-S}$. Greybody factors, entropy production, backreaction, and island contributions can modify the relation between mass and the effective subsystem dimensions.
A first-principles treatment would also require inverse spectral moments of the relevant microstate Gram matrices. Existing results in AdS, Minkowski, rotating, charged, and supersymmetric settings provide possible starting points~\cite{Penington:2019kki,Balasubramanian:2022gmo,
Balasubramanian:2022lnw,Climent:2024trz,Muck:2024fpb,Iizuka:2024njd}.
Finally, the singularity at $\alpha=1$ is a property of the unregularized inverse-Gram variance. Ridge regularization or truncation of near-zero Gram eigenvalues turns it into a finite crossover.
{Establishing a quantum phase transition would require a specified thermodynamic or dynamical limit and an analysis of finite-size scaling.}

\newpage
\appendix

\section{Marchenko-Pastur variance for the associated linear problem}

Consider the complex $N\times P$ linear reconstruction problem for one output channel~\cite{speicher2023hda},
\begin{equation}
\widetilde y=y_0+\epsilon,\qquad
\mathbb E[\epsilon\mid X]=0,\qquad
\mathbb E[\epsilon\epsilon^\dagger\mid X]=\sigma^2 I_N,
\qquad X\in\mathbb C^{N\times P}.
\end{equation}
This calculation separates the effect of additive target noise on the coefficients from the underlying noiseless least-squares or minimum-norm solution.
No realizability assumption $y_0\in\operatorname{Ran}X$ is required.
We use isotropic covariance only to simplify the variance into a scalar inverse-Gram moment.
For a general covariance the result depends on its overlap with the Gram eigendirections, as noted in Sec.~III.
For the ensemble average take $X=G/\sqrt N$, where $G$ has independent standard circular complex Gaussian entries, and let $P/N\to\alpha\ne1$.

For $P<N$, the noiseless least-squares solution is $w_<^{(0)}=(X^\dagger X)^{-1}X^\dagger y_0$, so
\begin{equation}
\delta w_<=(X^\dagger X)^{-1}X^\dagger\epsilon,
\qquad
\mathbb E_\epsilon\|\delta w_<\|^2
=\sigma^2\operatorname{Tr}(X^\dagger X)^{-1}.
\end{equation}
Thus the variance is controlled by the inverse positive eigenvalues of $X^\dagger X$. The complex-Wishart inverse moment gives
\begin{equation}
\frac{1}{N}\mathbb E_{X,\epsilon}\|\delta w_<\|^2
\longrightarrow\sigma^2\frac{\alpha}{1-\alpha},
\qquad 0<\alpha<1.
\end{equation}

For $P>N$, the noiseless minimum-norm interpolant is $w_>^{(0)}=X^\dagger(XX^\dagger)^{-1}y_0$ and
\begin{equation}
\delta w_>=X^\dagger(XX^\dagger)^{-1}\epsilon.
\end{equation}
Hence
\begin{equation}
\frac{1}{N}\mathbb E_{X,\epsilon}\|\delta w_>\|^2
=\frac{\sigma^2}{N}\mathbb E_X\operatorname{Tr}(XX^\dagger)^{-1}
\longrightarrow\frac{\sigma^2}{\alpha-1},
\qquad \alpha>1.
\end{equation}
The $N$ eigenvalues of $XX^\dagger$ are precisely the positive eigenvalues of $X^\dagger X$. Using the positive $N\times N$ Gram operator on the $\alpha>1$ side automatically avoids the zero-mode normalization ambiguity.

The exact Haar normalization of the purification does not change these fixed-aspect-ratio limits. Writing $T=\|G\|_F^2$ and $U=G/\sqrt T$, so that $X=\sqrt{T/N}\,U$ and $D=\sqrt P\,U$, one finds that the corresponding inverse-Gram averages differ only by a finite-size factor tending to one. Standard regularization, such as ridge regression or an explicit cutoff on small Gram eigenvalues, suppresses the near-zero modes and turns the unregularized interpolation pole into a finite crossover.

\section{Direct reconstruction from the purification Gram operators}

For the matrix approximation problem
\begin{equation}
\min_X\|Y-DX\|_F^2,
\end{equation}
variation with respect to $X^*$ gives the normal equation $D^\dagger(DX-Y)=0$.

For $P<N$ and generic full column rank,
\begin{equation}
X_<=(D^\dagger D)^{-1}D^\dagger Y,
\qquad
\Pi_B^\psi=D(D^\dagger D)^{-1}D^\dagger.
\end{equation}
The matrix $\Pi_B^\psi$ is Hermitian, idempotent, and satisfies $\Pi_B^\psi D=D$. Hence it is the orthogonal projector onto $\mathcal H_B^\psi=\operatorname{Col}(D)=\operatorname{supp}\rho_B$. The two terms in
\begin{equation}
Y-DX=(I_B-\Pi_B^\psi)Y+(\Pi_B^\psi Y-DX)
\end{equation}
lie in orthogonal black hole sectors. Therefore
\begin{equation}
\min_X\|Y-DX\|_F^2=\|(I_B-\Pi_B^\psi)Y\|_F^2.
\end{equation}
Applying this to $Y=W_BD$ gives the pre-Page minimum, and a zero-residual solution exists precisely when $(I_B-\Pi_B^\psi)W_B\Pi_B^\psi=0$.

For $P>N$ and generic full row rank,
\begin{equation}
X_>=D^\dagger(DD^\dagger)^{-1}Y,
\qquad DX_>=Y.
\end{equation}
Every exact solution has the form $X=X_>+Z$ with $DZ=0$. Since the columns of $X_>$ lie in $\operatorname{Ran}D^\dagger=(\ker D)^\perp$, they are orthogonal to the columns of $Z$. Hence $\|X_>+Z\|_F^2=\|X_>\|_F^2+\|Z\|_F^2$. Thus $X_>$ is the unique minimum-Frobenius-norm member of the exact-solution family.

The equality of the positive spectra of $D^\dagger D$ and $DD^\dagger$ is the standard singular-value relation: if $D^\dagger Dv=\lambda v$ with $\lambda>0$, then $Dv\ne0$ is an eigenvector of $DD^\dagger$ with the same eigenvalue, and conversely. Hence $D^\dagger D=P\rho_R^T$ and $DD^\dagger=P\rho_B$ have identical positive eigenvalues, with null sectors determined by the dimension imbalance {for generic full-rank $D$}.

Finally, suppose exact reconstruction $DX=W_BD$ holds with $W_B=W_B^\dagger$. If the radiation representative is also Hermitian, $X=X^\dagger$, then $DXD^\dagger$ is Hermitian and
\begin{equation}
W_BDD^\dagger=DD^\dagger W_B
\qquad\Longleftrightarrow\qquad
[W_B,\rho_B]=0.
\end{equation}
In the full-row-rank regime, the converse is also true. Since $[W_B,DD^\dagger]=0$ also implies $[W_B,(DD^\dagger)^{-1}]=0$, one can see that the operator $X=D^\dagger(DD^\dagger)^{-1}W_BD$ satisfies $X^\dagger=X$.
This gives the Hermiticity condition directly in terms of the physical black hole density operator.

  \bibliographystyle{elsarticle-num}

\begin{thebibliography}{10}
\expandafter\ifx\csname url\endcsname\relax
  \def\url#1{\texttt{#1}}\fi
\expandafter\ifx\csname urlprefix\endcsname\relax\def\urlprefix{URL }\fi
\expandafter\ifx\csname href\endcsname\relax
  \def\href#1#2{#2} \def\path#1{#1}\fi

\bibitem{Hawking1976}
S.~W. Hawking, Breakdown of predictability in gravitational collapse, Phys. Rev. D 14 (1976) 2460--2473.

\bibitem{Almheiri2021}
A.~Almheiri, T.~Hartman, J.~Maldacena, E.~Shaghoulian, A.~Tajdini, The entropy of hawking radiation, Rev. Mod. Phys. 93 (2021) 035002.

\bibitem{Page1993a}
D.~N. Page, Average entropy of a subsystem, Phys. Rev. Lett. 71 (1993) 1291--1294.

\bibitem{Page1993b}
D.~N. Page, Information in black hole radiation, Phys. Rev. Lett. 71 (1993) 3743--3746.

\bibitem{Akers2022}
C.~Akers, G.~Penington, Quantum minimal surfaces from quantum error correction, SciPost Phys. 12 (2022) 157.

\bibitem{Kar2023}
A.~Kar, Non-isometric quantum error correction in gravity, JHEP 02 (2023) 195.

\bibitem{Akers:2022qdl}
C.~Akers, N.~Engelhardt, D.~Harlow, G.~Penington, S.~Vardhan, The black hole interior from non-isometric codes and complexity, JHEP 06 (2024) 155.

\bibitem{Antonini:2024yif}
S.~Antonini, V.~Balasubramanian, N.~Bao, C.~Cao, W.~Chemissany, Non-isometry, state dependence and holography, JHEP 02 (2025) 150.

\bibitem{Belkin2019}
M.~Belkin, D.~Hsu, S.~Ma, S.~Mandal, Reconciling modern machine-learning practice and the classical bias-variance trade-off, Proc. Natl. Acad. Sci. USA 116~(32) (2019) 15849--15854.

\bibitem{Loog2020}
M.~Loog, T.~Viering, A.~Mey, J.~H. Krijthe, D.~M.~J. Tax, A brief prehistory of double descent, Proc. Natl. Acad. Sci. USA 117~(20) (2020) 10625--10626.

\bibitem{Schaeffer2023}
R.~Schaeffer, M.~Khona, Z.~Robertson, A.~Boopathy, K.~Pistunova, J.~W. Rocks, I.~R. Fiete, O.~Koyejo, Double descent demystified: Identifying, interpreting \& ablating the sources of a deep learning puzzle (2023).
\newblock \href {http://arxiv.org/abs/2303.14151} {\path{arXiv:2303.14151}}.

\bibitem{Marchenko1967}
V.~A. Marchenko, L.~A. Pastur, Distribution of eigenvalues for some sets of random matrices, Math. USSR-Sb. 1~(4) (1967) 457--483.

\bibitem{speicher2023hda}
R.~Speicher, High dimensional analysis: Random matrices and machine learning, Lecture notes, Saarland University, summer Term 2023 (2023).

\bibitem{Schuld:2018uel}
M.~Schuld, N.~Killoran, Quantum machine learning in feature hilbert spaces, Phys. Rev. Lett. 122 (2019) 040504.

\bibitem{Kempkes:2025hiw}
M.~Kempkes, A.~Ijaz, E.~Gil-Fuster, C.~Bravo-Prieto, J.~Spiegelberg, E.~van Nieuwenburg, V.~Dunjko, Double descent in quantum kernel methods, PRX Quantum 7 (2026) 010312.

\bibitem{Tomasi:2025wkm}
J.~Tomasi, S.~Anthoine, H.~Kadri, Benign overfitting with quantum kernels, in: E.~Perkovi{\'c}, D.~Malinsky (Eds.), Proceedings of the 42nd Conference on Uncertainty in Artificial Intelligence, Vol. 337 of Proceedings of Machine Learning Research, PMLR, 2026, pp. 6740--6766.

\bibitem{Kamisoyama:2026psl}
K.~Kamisoyama, L.~Nagano, K.~Terashi, Double descent in quantum kernel ridge regression (2026).
\newblock \href {http://arxiv.org/abs/2604.17202} {\path{arXiv:2604.17202}}.

\bibitem{Hawking1975}
S.~W. Hawking, Particle creation by black holes, Commun. Math. Phys. 43 (1975) 199--220, erratum: Commun. Math. Phys. 46 (1976) 206.

\bibitem{Penington:2019kki}
G.~Penington, S.~H. Shenker, D.~Stanford, Z.~Yang, Replica wormholes and the black hole interior, JHEP 03 (2022) 205.

\bibitem{Kawabata:2021hac}
K.~Kawabata, T.~Nishioka, Y.~Okuyama, K.~Watanabe, Probing hawking radiation through capacity of entanglement, JHEP 05 (2021) 062.

\bibitem{Balasubramanian:2022gmo}
V.~Balasubramanian, A.~Lawrence, J.~M. Magan, M.~Sasieta, Microscopic origin of the entropy of black holes in general relativity, Phys. Rev. X 14 (2024) 011024.

\bibitem{Balasubramanian:2022lnw}
V.~Balasubramanian, A.~Lawrence, J.~M. Magan, M.~Sasieta, Microscopic origin of the entropy of astrophysical black holes, Phys. Rev. Lett. 132 (2024) 141501.

\bibitem{Climent:2024trz}
A.~Climent, R.~Emparan, J.~M. Magan, M.~Sasieta, A.~Vilar~Lopez, Universal construction of black hole microstates, Phys. Rev. D 109 (2024) 086024.

\bibitem{Muck:2024fpb}
W.~M{\"u}ck, Black holes and marchenko-pastur distribution, Phys. Rev. D 109 (2024) 126001.

\bibitem{Iizuka:2024njd}
N.~Iizuka, M.~Nishida, A note on the non-planar corrections for the page curve in the pssy model via the iop matrix model correspondence, JHEP 12 (2024) 212.

\end{thebibliography}


\end{document}